\documentclass[fleqn,usenatbib]{mnras}

\usepackage{newtxtext,newtxmath}
\usepackage{hyperref}
\usepackage{xcolor}
\hypersetup{
  colorlinks   = true, 
  urlcolor     = black, 
  linkcolor    = blue, 
  citecolor   = blue 
}

\usepackage[T1]{fontenc}

\DeclareRobustCommand{\VAN}[3]{#2}
\let\VANthebibliography\thebibliography
\def\thebibliography{\DeclareRobustCommand{\VAN}[3]{##3}\VANthebibliography}

\usepackage{graphicx}	
\usepackage{amsmath}	
\usepackage{amssymb}	

\title[LOFAR discovery of radio relic]{Discovering the most elusive radio relic in the sky: Diffuse Shock Acceleration caught in the act?}

\author[N.~T.~Locatelli et al.]
{
\parbox{\textwidth}{
Nicola~T.~Locatelli$^{1,2}$, 
Kamlesh Rajpurohit$^{1}$, 
Franco Vazza$^{1,2,3}$,
Fabio Gastaldello$^{4}$,
Daniele Dallacasa$^{1,2}$,
Annalisa Bonafede$^{1,2}$, 
Mariachiara Rossetti$^{4}$,
Chiara Stuardi$^{1,2}$,  
Etienne Bonassieux$^{1}$,
Gianfranco Brunetti$^{2}$,
Marcus Br\"{u}ggen$^{4}$,
Timothy Shimwell$^{5,6}$ 
}
\vspace{0.4cm} \\
\parbox{\textwidth}{
$^1$Università di Bologna, Dip. di Fisica \& Astronomia DIFA, via Gobetti 92, Bologna, Italy \\
$^2$INAF -- Istituto di Radioastronomia, via Gobetti 92, Bologna, Italy \\
$^3$University of Hamburg, Hamburger Sternwarte, Gojenbergsweg 112, 21029 Hamburg, Germany \\
$^4$INAF – IASF Milano, Via Bassini 15, 20133 Milano, Italy \\
$^5$Leiden  Observatory,  Leiden  University,  PO  Box  9513,  2300  RALeiden, The Netherlands \\
$^6$ASTRON, The Netherlands Institute for Radio Astronomy, Postbus2, 7990 AA Dwingeloo, The Netherlands 
}
}

\date{Accepted 2020 April 20. Received 2020 April 20; in original form 2020 March 19}

\pubyear{2020}

\begin{document}
\label{firstpage}
\pagerange{\pageref{firstpage}--\pageref{lastpage}}
\maketitle

\begin{abstract}
The origin of radio relics is usually explained via diffusive shock acceleration (DSA) or re-acceleration of electrons at/from merger shocks in galaxy clusters. The case of acceleration is challenged by
the low predicted efficiency of low-Mach number merger shocks, unable to explain the power observed in most radio relics.
In this Letter we present the discovery of a new giant radio relic around the galaxy cluster Abell 2249 ($z=0.0838$) using LOFAR.  It is special since it has the lowest surface brightness of all known radio relics. We study its 
radio and X-ray properties combinig LOFAR data with uGMRT, JVLA and XMM.
This object has a total 
power of  $L_{1.4\rm GHz}=4.1\pm 0.8 \times 10^{23}$ W Hz$^{-1}$
and integrated spectral index $\alpha = 1.15\pm 0.23$. 
We infer for this radio relic a lower bound on the magnetisation of $B\geq 0.4\, \mu$G, a shock Mach number of $\mathcal{M}\approx 3.79$,  and a low acceleration efficiency consistent with DSA. 
This result suggests that a missing population of relics may become visible thanks to the unprecedented sensitivity of the new generation of radio telescopes.
\end{abstract}




\section{Introduction}
Radio relics are elongated, arc-shaped diffuse synchrotron sources extened over $\sim$ Mpc, usually found at the periphery of clusters of galaxies with ongoing mergers, showing with steep spectrum ($\alpha>1$, where the flux density $S_\nu$ is defined as $S_\nu\propto\nu^{-\alpha}$ and $\alpha$ is the spectral index) steepening towards the cluster centre \citep[e.g.][for a review]{vanweeren19}.
Radio relics are strongly polarized at high frequencies, with a polarization fraction that can go up to $20-30\%$ at 1.4~GHz and $\sim 70\%$ at 5~GHz \citep{vanweeren10,kierdorf17,loi17}.
Several radio relics have also been found to trace the position of shock waves, as detected as discontinuities in the X-ray brightness profiles of the intra-cluster medium (ICM)
\citep{2013PASJ...65...16A, 2018MNRAS.476.5591B}. 
Merger shock waves are believed to be generated when clusters of galaxies collide, and then propagate along the direction of the merger. Shocks are more easily seen edge-on as projection boosts their surface brightness, and the same observational bias should also apply to radio relics. 
The kinetic energy dissipated at shocks 
should be related to the powering of the radio emission, via Diffusive Shock Acceleration (DSA, \citealt{bell78I,jones91}), as
originally proposed by \citet{1998A&A...332..395E}. However, the Mach numbers that are independently inferred from discontinuities observed in X-rays are generally too weak ($\mathcal{M} \sim 2$) to account for the required electron acceleration efficiency by DSA in relics \citep[e.g.][hereafter B+20]{2020A&A...634A..64B}.
Moreover, shock waves in the intracluster medium should also accelerate protons that would create $\gamma$-ray emission in the collision with the thermal protons of the ICM. These $\gamma$-rays have not been detected \citep{ackermann16}, which translates into 
limits on the maximum acceleration efficiency of protons in structure formation shocks ($<10^{-3}$, \citealt{vazza16}).
This conundrum can be by-passed when invoking a pre-existing population of mildly non-thermal electrons that get re-accelerated by the shocks \citep{pinzke13,kang-ryu15, 2005ApJ...627..733M}. In a few cases, Active Galactic Nuclei  could have supplied the relativistic electrons in the upstream region of the shock that creates the relic \citep{bonafede14,vanweeren17,stuardi19}.
Both acceleration and reacceleration processes operate in the ICM and should contribute to the population of radio relics.
We note that we adopted a flat-$\Lambda$CDM cosmology with $H_0=69.6\rm \,km\,s^{-1}\,Mpc^{-1}$ and $\Omega_M=0.286$ throughout the paper.

\vspace{-0.3cm}
\subsection{General properties of Abell 2249}

In this work we present the discovery of a giant radio relic found at the periphery of the galaxy cluster Abell\,2249 (hereafter A2249; RA 257.44080, DEC 34.45566). 
Its galaxies and ICM features have been studied in details at various wavelengths by a number of authors:
the cluster mean redshift is z=0.0838 \citep{lagana19,lopes18,bulbul16}; 
the velocity dispersion of its constituent galaxies is between $\sigma_{\rm vel}=894\pm50$ \citep{lopes18} and $976\pm38$~km\,s$^{-1}$ \citep{oh18}.
\citealt{lagana19} provided detailed XMM-Newton maps of temperature (peaking in the 4-7 keV energy band), pseudo-pressure, pseudo-entropy and metallicity in the central region, within the first $\sim 400$~kpc from the cluster centre. 
They classified A2249 as a non-cool-core (NCC) disturbed cluster, although \citealt{oh18} and \citealt{lopes18} did not find evidence for merging from the spectroscopic redshift distribution of cluster members.  However, a Dressler \& Shectman three-dimensional test of the galaxy redshifts  suggests that the cluster is disturbed  \citep{lopes18}.
The radius of the cluster is R$_{500}=1.56\pm 0.06,\,1.1^{+0.3}_{-0.1}$~Mpc depending on the mass estimate which is debated between the values M$_{500}=11.7\pm 1.4,\, 4.0^{+5.2}_{-1.1},\, 3.73^{+0.18}_{-0.19} $, in units of $10^{14}~M_\odot$, derived from radial velocity distribution, Chandra and Planck data respectively \citep{lopes18, zhu16, 2016A&A...594A..27P}. In this article we will adopt the Planck value.
At larger radii  R$_{200}=2.2\pm 0.1$~Mpc and M$_{200}=12.7\pm 1.5\times 10^{14} M_\odot$ \citep{lopes18, oh18}. 

\begin{figure*}
	 \includegraphics[width=0.7\textwidth]{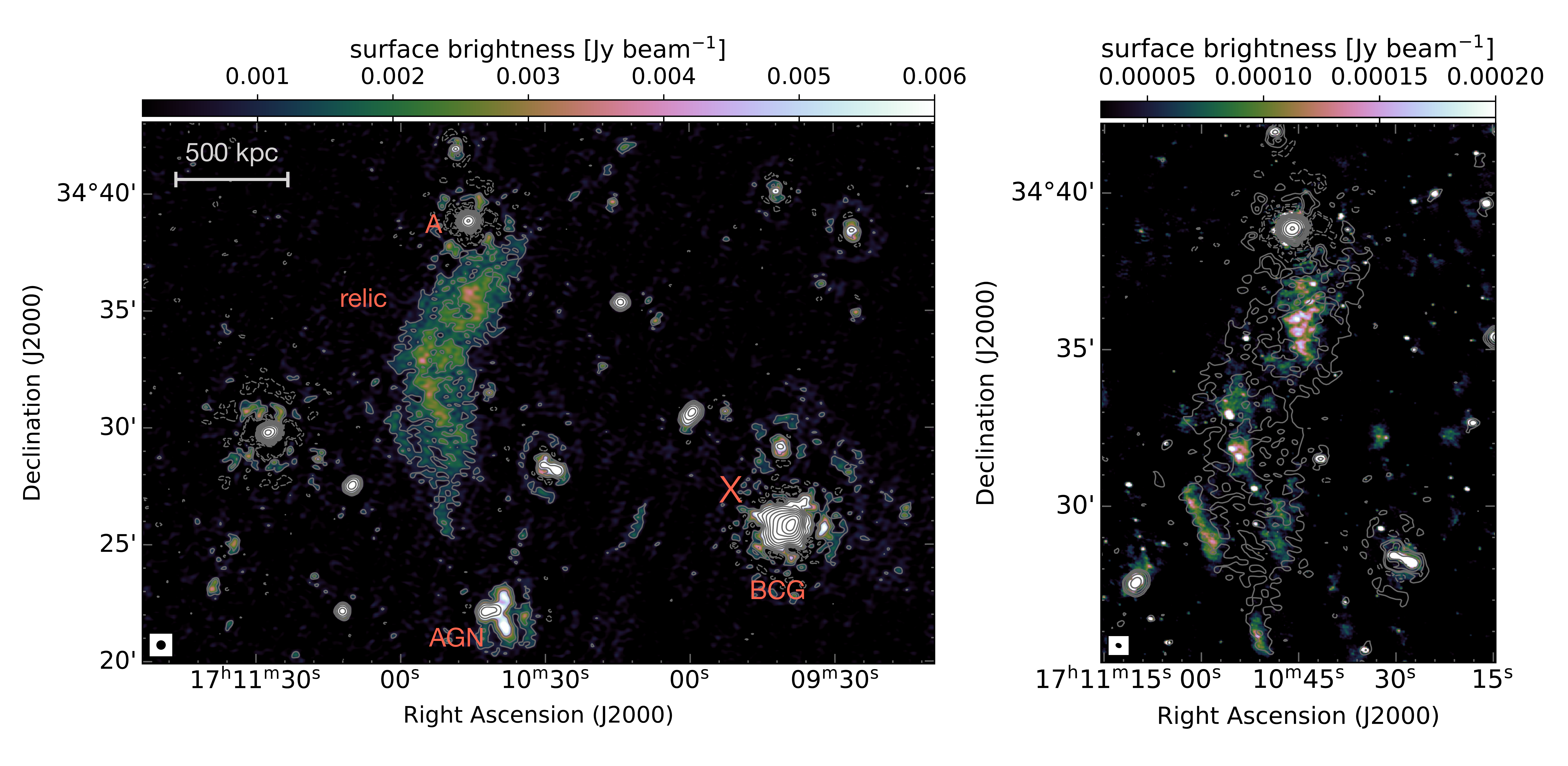}
      \includegraphics[width=0.29\textwidth]{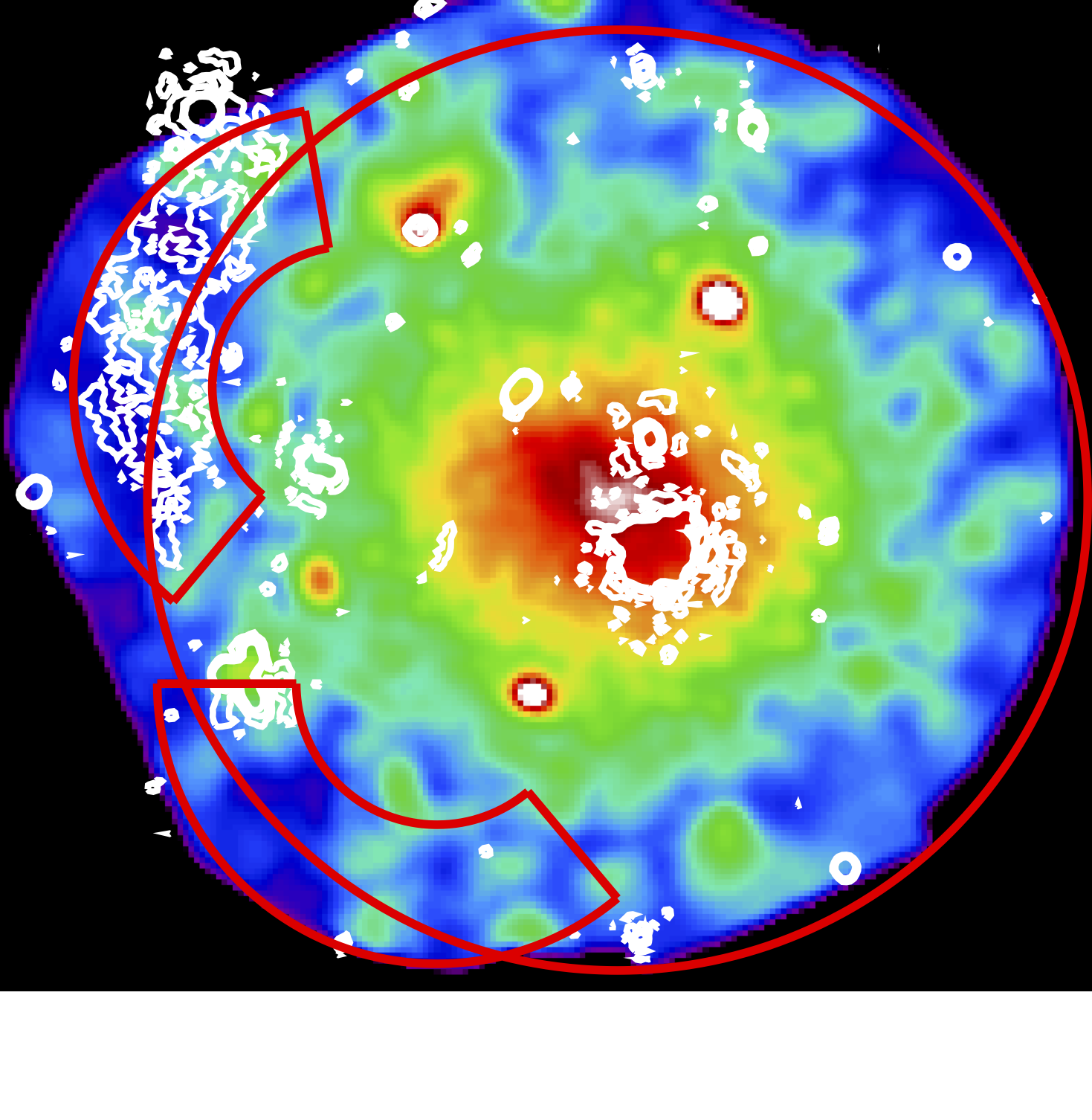}
    \vspace{-0.1cm}
    \caption{{\it Left:} LOFAR 
    low resolution ($20\arcsec$) image
    of Abell 2249, showing a spectacular large-scale radio relic. The red cross marks the cluster center. 
    Contour levels are drawn at $[1,2,4,8,\dots]\,\times\,3\,\sigma_{{\rm{ rms}}}$ and are from the LOFAR image. Negative $-3\sigma_{\rm rms}$ contours are shown with dotted lines
    {\it Centre:} uGMRT high resolution ($8\arcsec\times6\arcsec$) image of the relic, overlaid with LOFAR contours, revealing filamentary substructures.
    {\it Right:} Background subtracted, exposure corrected and adaptively smoothed XMM image in the 0.7-1.2 keV band of A2249. 
    The 144 MHz contours at 3,6,10$\sigma$ of the low resolution LOFAR radio emission are overlaid in white.
    A circle of radius $14\arcmin$ is drawn to guide the eye for the two sectors used in the spectral analysis described in the text: one encompassing the relic radio emission and one test region of the same extension at the same radial distance from the cluster centre.}
    \label{fig:LOFAR}
\end{figure*}
%

\section{Observation and data reduction}

\subsection{Radio observations}
The low frequency observations of the A2249 field was carried out with the LOw Frequency ARray (LOFAR). The LOFAR HBA ($120-168$MHz) observation was carried out during Cycle 9 (Proposal Id:LC9\_020). The centre of the pointing was not at the cluster centre, but at coordinates 17:01:13 +33:20:15 (RA, DEC), at a distance of 2.1 degrees. The on-source time is 8 hr with two scans of 10 min each on the flux calibrator 3C295. A first calibration and imaging run was performed 
using the LOFAR data reduction pipeline 
(v2.2\footnote{https://github.com/mhardcastle/ddf-pipeline}) involving both direction-independent \citep{2019A&A...622A...5D} and -dependent calibration of the data
\citep{2017A&A...598A.104S}.
Exploiting the sky models derived from the pipeline, we subtracted from the uv-data all sources outside a 1.9$^{\circ}\times1.9^{\circ}$ region centred on the relic. This was done using the PYthon Blob Detector and Source Finder (pybdsf; \citealt{2015ascl.soft02007M}). The resulting data was then self-calibrated through nine iteration steps and then imaged using WSClean v2.4 \citep{2014MNRAS.444..606O}.

We produced images at $6\arcsec$ (Fig.~\ref{fig:LOFAR}, left panel) and $20\arcsec$ resolution using a Briggs weighting scheme with robust -0.5. 
The image at higher (lower) resolution has a rms noise floor of $230 (350)\,\mu \rm Jy beam^{-1}$.
We determined and applied a correction factor \citep[see also]{2016ApJS..223....2V, 2016MNRAS.462.1910H} to match the LOFAR HBA flux densities of point-like sources with the ones derived from the TIFR GMRT Sky Survey (TGSS; \citealt{2017A&A...598A..78I}). We assume flux uncertainties of 20\% , similar to the LOFAR Two-meter Sky Survey images \citep{2019A&A...622A...1S}.\\

We also observed the cluster with the upgraded Giant Meter Radio Telescope (uGMRT), in Band-4 covering a frequency range of 550-950\,MHz (proposal DDT-C100). The data were flagged and calibrated using CASA. We then ran several rounds of direction-dependent self-calibration using the LOFAR {\tt DDF-pipeline} (see above). The image reaches a noise level of $16\,\mu \rm Jy beam^{-1}$ at 700\,MHz.\\

We have also analysed two short snapshot observations at 1.46 GHz from the VLA archive. About 8 min (four 2-min scans well spaced in time) and 25 min (single scan) in C and D configuration 
were available (project codes AS220 and AG294, respectively). We obtained a combined image of the intersecting part of the bands after standard calibration of the two individual datasets. The pointing was set on the brightest central galaxy (BCG), which is about 15$\arcmin$ off the relic position. This highly affected the local sensitivity. The combined C+D image (not shown) allowed a resolution of about 30$\arcsec$ and presents a number of separate patches of diffuse emission with peaks just above the local 3$\sigma$ in the region of the relic, with roughly the same morphology of the uGMRT image.

\subsection{X-ray: XMM observation}
A2249 (also known under the name PSZ2 G057.61+34.93) has been observed as part of the XMM Heritage Cluster Project\footnote{http://xmm-heritage.oas.inaf.it}, a large and unbiased sample of 118 clusters, detected with a high signal-to-noise ratio in the Second Planck SZ Catalogue.  We reduced the data with SAS v $16.1$. The observation with OBSID 0827010501 has a total clean exposure time of $20.4$ ks with MOS1, $20.7$ with MOS2 and $16.1$ with pn after filtering for soft proton flares (81\% of the total time for MOS and 93\% for the pn). We estimated the amount of residual soft protons following the procedure described in \citet{2019A&A...628A..83C} and found it to be negligible. For a full description of data reduction, image production and spectral extraction we refer to \citet{2019A&A...621A..41G}. In the right panel of Fig.~\ref{fig:LOFAR} we show the XMM image in the 0.7-1.2 keV band with the overlay of the radio contours at 150 MHz with $6\arcsec$ resolution and the regions used for the spectral analysis.
Given that the emission of the cluster is filling the entire field of view of XMM for the estimate of the sky background components in a similar way to \citet{Snowden.ea:08} we used a spectrum from the ROSAT All-Sky Survey extracted from an anulus between 0.5 and 1 degree from the source. We fixed the Galactic $\rm{N_H}$ to $2.38\times10^{20}$ cm$^{-2}$ at HI LAB value \citep{Kalberla.ea:05} given the negligible difference with the value ($2.5\times10^{22}$ cm$^{-2}$) which estimates the possible contribution of molecular hydrogen \citep{Willingale.ea:13}.
\begin{table}
\caption{Properties of the Cornetto relic in Abell 2249 (z=0.0838).}
\centering
\begin{tabular}{l c c}
    Symbol & value & description \\
\hline\hline
    $F_{144\rm MHz}$  & $370\pm 70\,\rm mJy$ & flux density at 144\,MHz\\
    $L_{144\rm MHz}$  & $5.9\pm 1.2\rm \times 10^{24}\, W \,Hz^{-1}$ & luminosity at 144\,MHz\\
    $F_{700\rm MHz}$  & $60\pm 12\,\rm mJy$ & flux density at 700\,MHz\\
    $\alpha_{700\rm MHz, \, int}^{144\rm MHz}$ & $1.15\pm 0.23$ & integrated spectral index \\
    & & between 144 and 700\,MHz\\
    $\Delta\Omega_{144\rm MHz} $&$ 28.46 \, \rm arcmin^2$ & relic solid angle at 144\,MHz\\
    $R_{\rm proj}$ & $1.40\rm \,Mpc$ & projected radial cluster distance \\
    LAS & 13.2$\arcmin$ & largest angular scale\\
    LLS(z$_{\rm A2249}$) & 1.3 Mpc & largest linear scale \\
\hline
\end{tabular}
\label{tab:parameters}
\end{table}
%
\section{Results}
\paragraph*{Morphology}

The extended diffuse emission at 144\,MHz (Fig.~\ref{fig:LOFAR}) is arc-shaped and oriented perpendicular to the radial direction from the cluster centre, in the North-East-East sector of A2249, spanning 
an angular radial range $[11.0; 17.0] \arcmin$ from the cluster centre.
The relic width  is maximal at its the mean azimuthal direction and is minimal at the azimuthal ends of the diffuse emission, giving the radio relic a shape very similar to a crescent moon or the popular Italian sweet bun named "cornetto". 
The brightest part of the relic at 144\,MHz is found at an angular radial distance of $\simeq 14.7 \arcmin$, that is a linear distance of 1.40~Mpc at the redshift of A2249.
The relic's largest angular scale (LAS) is $\simeq 13.2 \arcmin$, corresponding to a physical size of 1.3~Mpc at the  redshift of the cluster. 
The northern end of the diffuse emission coincides with a bright 
unresolved radio source (A, Fig.~\ref{fig:LOFAR} left panel), of $400\, \rm mJy beam^{-1}$ at 144 MHz.
Visible in the south-west direction is the BCG of A2249.
Deconvolution artefacts remained around the bright sources A and BCG. 
The relic also shows elongated patches of emission of a few arcminutes, in analogy with the filamentary structures described in other radio relics \citep{2014ApJ...794...24O, pearce17,rajpurohit18}, whose origin is still unclear.
The image at 700\,MHz also shows diffuse emission at the relic position above $3\sigma$, with a similar morphology as at lower frequency, as well as a large density of point sources (Fig.~\ref{fig:LOFAR} central panel).

\paragraph*{Radio spectrum \& luminosity}
The flux density and luminosity of the Cornetto relic at 144~MHz are $F_{144\rm MHz}=370\pm 70\,\rm mJy$ and $L_{144\,\rm MHz} =  5.9\pm1.2 \times 10^{24} \rm W \, Hz^{-1}$, respectively. 
The integrated spectral index, calculated from the ratio of the total flux densities at 144 and 700 MHz in the relic region (determined at 144 MHz) is $\alpha=1.15\pm0.23$. The observed quantities are summarized in Tab.~\ref{tab:parameters}.

\begin{figure}
    \centering
    \includegraphics[width=0.40\textwidth, 
    height=0.27\textwidth]{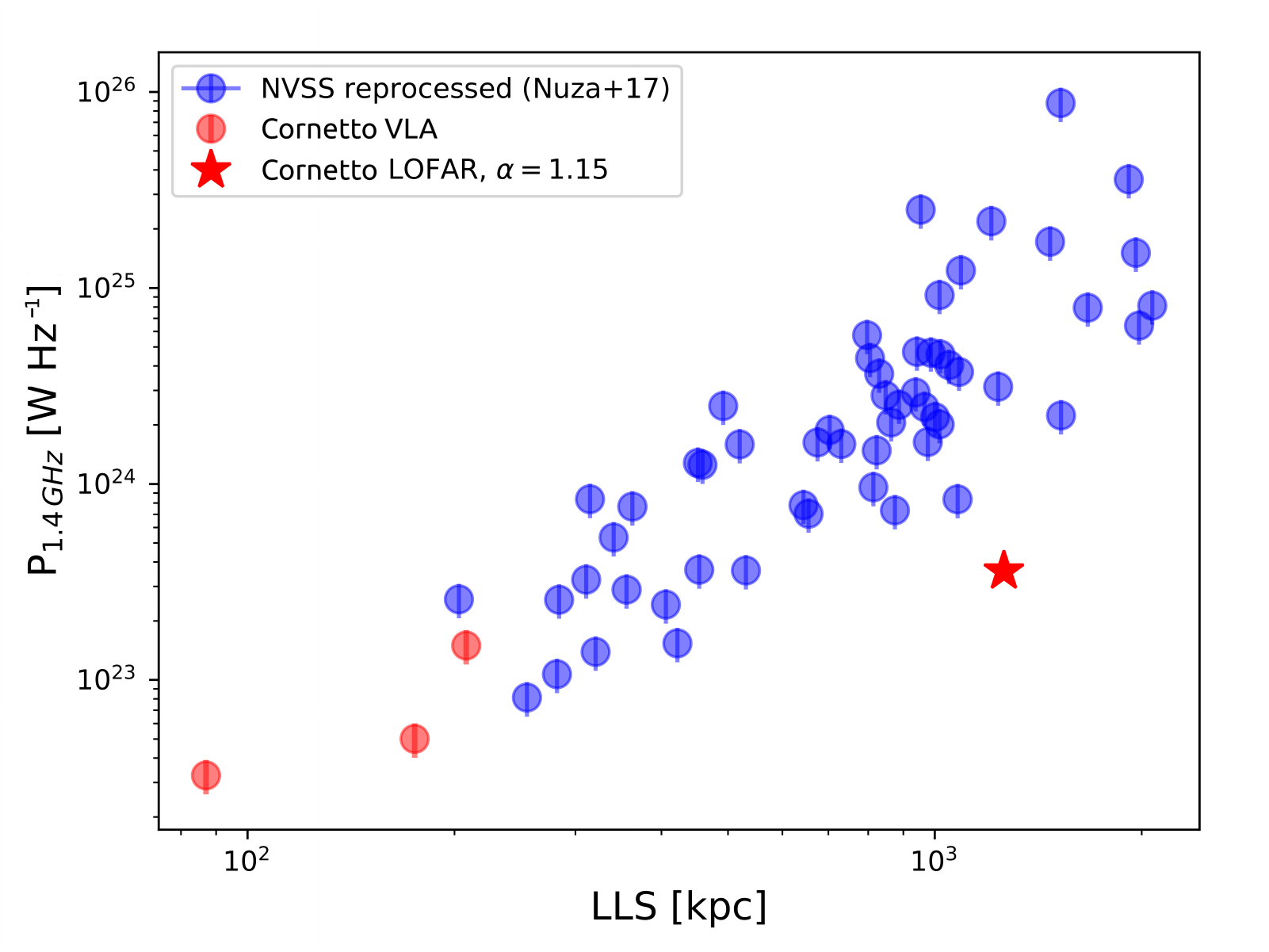} %
    \vspace{-0.4cm}
    \caption{The luminosity at 1.4 GHz is plotted against the LLS for the radio relics detected in the NVSS \citep{nuza17}. The red star shows the power of the Cornetto relic extrapolated to 1.4 GHz.}
    \label{fig:PradioVSLLS}
\end{figure}

Assuming $\alpha=1.15$ to be constant we extrapolated the luminosity at 1.4~GHz to be $L_{1.4\,\rm GHz} = 4.1\pm0.8 \times 10^{23}$ W~Hz$^{-1}$.
The Cornetto relic (red star, Fig.~\ref{fig:PradioVSLLS}) is found to lie below the observed scaling relation between the radio power at 1.4~GHz and the largest linear size (LLS) of a sample of know radio relics presented in \cite{nuza17}, extracted from the NRAO VLA Sky Survey (NVSS, \citealt{1998AJ....115.1693C}).
From  archival VLA images we find three different regions across the relic with matching $3\sigma$ contours between 144 MHz and 1.4 GHz. 
We computed the integrated power for these three regions and plotted them in Fig.~\ref{fig:PradioVSLLS} (red circles). 
The correlation in Fig.~\ref{fig:PradioVSLLS} has already been shown to be determined largely by the NVSS sensitivity \citep{nuza17}.
The LOFAR observations presented here seem to open the window to a population of faint and diffuse relics that have not been seen to date. 

\begin{figure}
    \centering
    \includegraphics[
    height=0.26\textwidth]
    {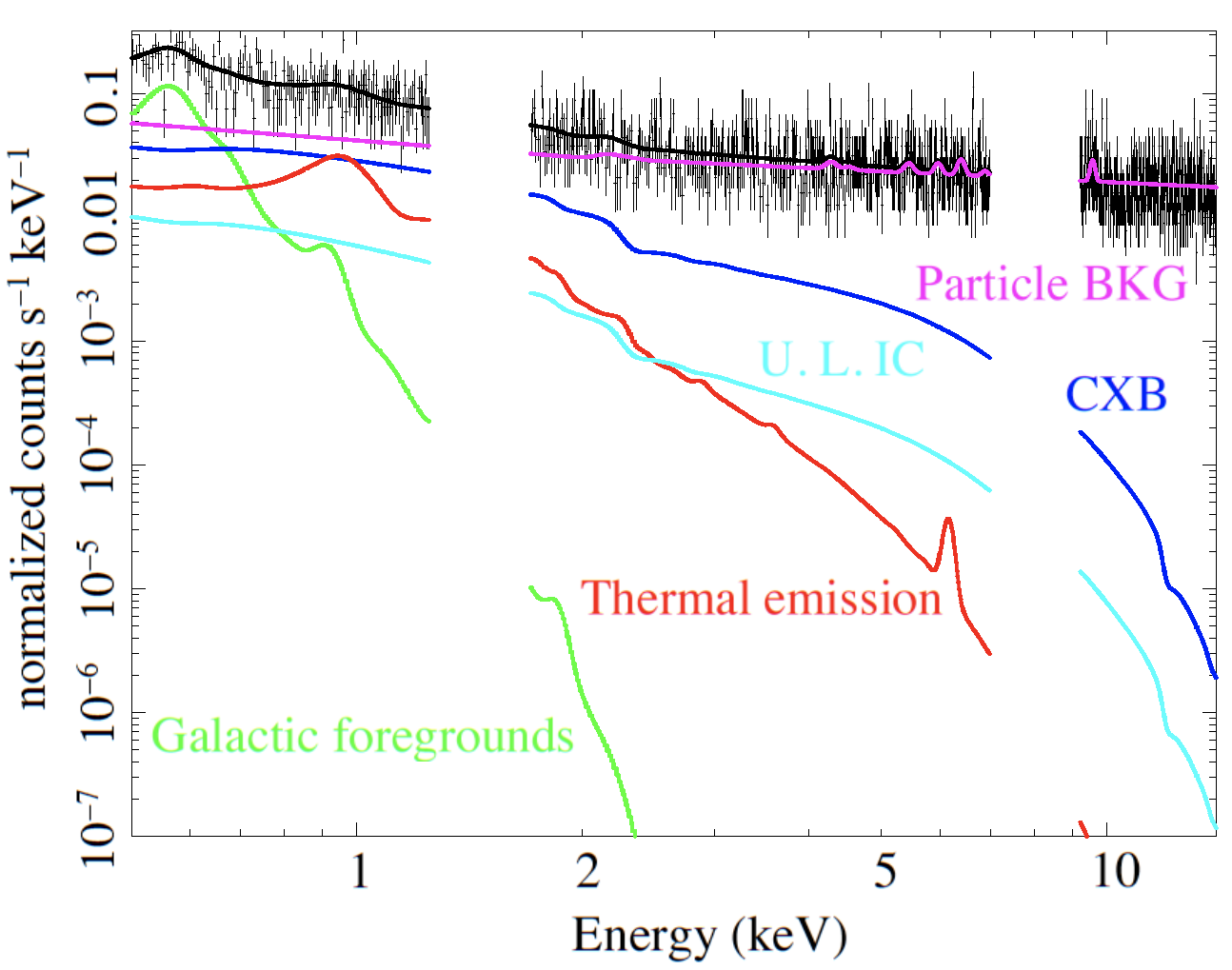}
   \vspace{-0.1cm} 
    \caption{XMM pn spectrum extracted from the region of the relic radio emission.  The magenta line shows the instrumental background, the green one the galactic foregrounds, the  blue one the Cosmic X-ray Background, the red line the ICM thermal emission and the  cyan one the 90\% upper limit on the IC power law.}
    \label{fig:Xspectra}
\end{figure}

\paragraph*{X-ray properties at the position of the relic}
We extracted XMM MOS and pn spectra from an angular sector which covers the relic radio emission as shown in Fig.~\ref{fig:LOFAR}, right panel.
The region extends beyond R$_{500}$ and therefore the thermal emission is below the background. The temperature obtained is prone to large systematic errors and we therefore rely on the value obtained within the full annulus of $kT=3.0\pm1.3$ keV together with an electron density $n_e=6.4\pm 1.5\times 10^{-4} \rm cm^{-3}$. Assuming that temperature 
we modeled the expected IC emission as a power law with fixed photon index of 2.15 as derived from the radio spectral index and extrapolated a 90\% upper limit of $1.0\times10^{-13}$ erg cm$^{-2}$ s$^{-1}$ in the 20-80 keV range. The X-ray spectrum in the relic region and its modeling is shown in Fig.~\ref{fig:Xspectra}. 
It is equivalent to the spectrum extracted from a region at the same radial distance from the cluster but avoiding the relic emission (see Fig.~\ref{fig:LOFAR}), confirming that any IC emission is clearly subdominant. 

\section{Modelling of physical properties} \label{sec:models}

Based on our  observations we study the origin of the relic in A2249 and infer limits on its magnetic field.

\paragraph*{Diffusive Shock Acceleration}

Assuming DSA, the power emitted by the Cornetto relic can be related to its shock properties \citep[e.g.][hereafter HB08; B+20]{hoeft08}:
\begin{equation}
    L_{\nu, \, \rm obs}= C \cdot \frac{A}{\rm Mpc^2} \cdot \frac{n_{e,d}}{10^{-4}\rm cm^{-3}} \cdot \xi_{e} \cdot  \frac{T_{e,d}^{3/2}}{\nu^{\alpha/2}} \frac{B^{1+\frac{\alpha}{2}}}{B^2+B^2_{\rm CMB}(z)} ,
    \label{eq:hb} 
\end{equation}
where $A$ is the surface area of the relic, calculated as LLS$~\cdot~d_{\rm thick}$ (we assume again $d_{\rm thick}=$LLS).  $n_{e,d}$ is the downstream electron density, $\xi_{e}$ is the (yet unknown) fixed fraction of the kinetic energy flux $\Phi_e/\Phi_k$ injected at the shock front into suprathermal electrons,
$T_{e,d}$ is the downstream electron temperature and $B_{\rm CMB}$ is the equivalent field of the Cosmic Microwave Background evaluated at the redshift of A2249. The normalisation $C$ equals to $6.4 \times 10^{34} \frac{\rm erg}{\rm s\,Hz}$ when $T_{e,d}$ in units of [7 keV k$_B^{-1}$], $\nu$ in units of 1.4 GHz and $B$ in [$\mu$G].

\begin{figure}
    \centering
    \includegraphics[width=0.45\textwidth]
    {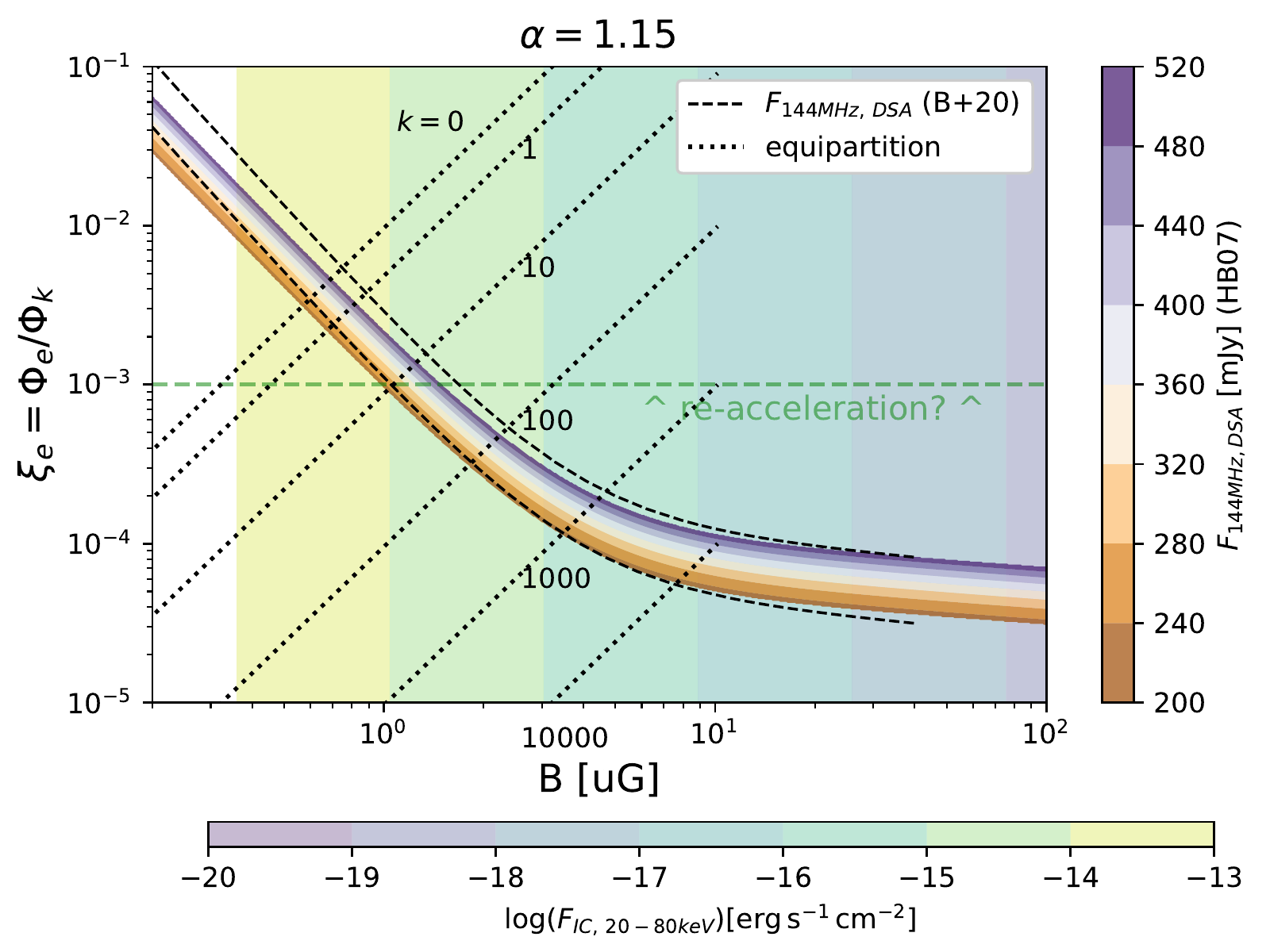}
    \vspace{-0.4cm}
    \caption{The ($B,\, \xi_e$) parameter space assuming $\alpha=1.15$.  The curves show the points that reproduce the $F_{144\rm MHz}$ within $2\sigma$ uncertainty 
    assuming DSA using HB08 (orange-white-purple) or B+20 (dashed black) formalism. The green-violet shaded background shows the inverse Compton fluxes expected in the $20-80$keV band. 
    The dotted lines show the values obtained assuming equipartition for different values of k.
    }
    \label{fig:param_space}
\end{figure}

Considering the values in Tab.~\ref{tab:parameters}, an integrated spectral index $\alpha=1.15$ (holding a Mach number $\mathcal{M}=\sqrt{(\alpha+1)/(\alpha-1)}=3.79$)
and the quantities derived from the XMM-Newton observations $k_B T_e\simeq 3.0\pm 1.3$~keV and $n_{e,d}=6.4\pm 1.5 \times 10^{-4} \rm cm^{-3}$, we can constrain the ($B,\,\xi_e$) parameter space to reproduce $F_{144\,\rm MHz}$ ({\it DSA} curves in Fig.~\ref{fig:param_space}).
For completenss, we also consider the formulation of the model as found in B+20, which enforces the relativistic invariance in the HB08 model, which is particularly relevant for weak shocks.
We obtain a magnetic field of $B=1.2\,\mu$G for $\xi_e=10^{-3}$ and $B=6.0\,\mu$G for $\xi_e=10^{-4}$.
The values for $\xi_e$ agree with models for DSA from shocks with Mach numbers $\mathcal{M}=3.5-4.0$ \citep[][]{kang-ryu15}. 
Larger efficiencies are hard to reconcile with DSA and (in other objects) are used to argue for the existence of a pre-existing electron population that may have been re-accelerated by an earlier episode of shock acceleration. 
Re-acceleration has  been invoked for most radio relics (all observed at frequencies $>600$~MHz) for which an underlying shock wave has been detected in X-rays at their location, with the exception is the radio relic in the El Gordo galaxy cluster (B+20). 
Instead, the efficiency required to power the Cornetto relic can be explained by DSA electrons from the thermal pool, by $\sim $ a few $\mu$G magnetic fields.

\paragraph*{Equipartition}
Synchrotron radiation provides information on both the electron's energy distribution and the magnetic field strength, $B$, in the medium.
A simplistic assumption 
to disentangle the
contribution of relativistic cosmic-rays (CRs) from magnetic fields is to assume equipartition between their energy densities in the plasma $\epsilon_{CR}=\epsilon_B$ \citep[e.g.][]{1997A&A...325..898B, 2005AN....326..414B}.
In this case, the total energy density of magnetic fields and of CRs $\epsilon_B+\epsilon_{CR}$ also approaches a minimum value. 
Classical equipartition formulae use parameters of the spectral energy distribution of electrons that is not affected by energy losses. In the case of radio relics instead, the spectrum of electrons emitting downstream results from the combination injection, transport and energy losses. We thus derive equipartition conditions assuming that the magnetic field in radio relics gets the same energy density of particles downstream, that is :
\begin{equation}
    \begin{split}
    \frac{1}{2}\rho_u\, \frac{\rm v^3_u}{\rm v_d}\, \xi_e\, (1+k) = \frac{B^2}{8\pi}
    \end{split}
\end{equation}
where $k$ is the ratio of energy budget between $p$ and $e$, $\rho$ and $\rm v$ are the gas density and shock velocity computed for the media respectively upstream (${}_u$) and downstream (${}_d$) of the shock front. The jump conditions have been derived from the shock Mach number $\mathcal{M}=3.79$.
With this approach $\xi_e$ is directly comparable with the values derived from DSA.

The results for $B,\, k$ and $\xi_e$ are degenerate, however the equipartition assumption alone constraints the parameter space between the curves for $k=0$ (indicating a plasma where the energy budget is only given by $e$) and $B<< 10\mu$G resulting from $\xi_e(1+k)<<1$. Combined with equipartition argument the efficiency selects the value of k.

\paragraph*{Inverse Compton scattering}
Based on the observed radio flux and assuming a power-law distribution of relativistic electrons, we can estimate the hard X-ray emission from Inverse Compton (IC) scatter from the same electron population responsible for the observed radio emission \citep[e.g.][]{govoni04}. 
Then we can compare this to recent upper limits obtained using XMM-Newton observations in the 0.1-12 keV band. 
We quote the flux estimates extrapolated in the 20-80 keV band for ease of comparison with previous estimates \citep[e.g.][]{2019A&A...628A..83C}. 
The IC flux $90\%$ upper limit $F_{\rm IC} \leq 1 \cdot 10^{-13} \rm erg\, cm^{-2}\, s^{-1}$ 
extrapolated in the $20-80$keV band sets a lower limit on $B>0.4 \mu$G.
A magnetic field strength of $B_{\rm low}=0.6~\rm \mu G$ (as suggested above assuming $\xi_e=10^{-3}$) or lower would result in IC emission larger than the  $F_{\rm IC} \approx 3.17 \cdot 10^{-14} \rm erg\, cm^{-2}\, s^{-1}$ upper limit derived for A523 by  \citet{2019A&A...628A..83C}. 
For comparison, $\alpha=1.15$ and $B=6.0 \rm \mu G$ (implying $\xi_e=10^{-4}$ for DSA) produces $F_{\rm IC} \approx 5 \cdot 10^{-15} \rm erg\, cm^{-2}\, s^{-1}$, i.e. about one order of magnitudes below present- day upper limits. 
The lower limit from IC combined with the limit $B << 10\mu$G from energy arguments implies efficiencies $\xi_e \in [5\cdot 10^{-5}-10^{-2}]$.
Larger values would violate equipartition.

\vspace{-0.3cm}
\section{Conclusions}

In this Letter we presented the discovery of extended, diffuse radio relic in A2249, found at low frequencies (120-168 MHz) with LOFAR. 
We have also observed the new relic (called Cornetto relic) at 700 MHz with the uGMRT and found patches of emission in coincidence of the brightest parts of the relic also in VLA archival data at 1.4 GHz. 
The magnetic field at the relic is estimated to be $B> 0.4 \mu$G, depending on model assumptions and the electron acceleration efficiency $\xi_e\leq 10^{-2}$ of the putative merger shock. 
The limits have been set from the absence of Inverse Compton emission in the $[0.1-12]$~keV energy band. 

The Cornetto relic is among the largest relics discovered to-date (13.2', corresponding to 1.26~Mpc) as well as the faintest one with such extent, once extrapolated at 1.4~GHz, lying at least a factor $\sim 10$ below the observed scaling relation between the radio power at 1.4~GHz and the LLS of radio relics. 

Its low luminosity is well explained by DSA for the inferred plasma and shock parameters, unlike most other radio relics that require a higher electron acceleration efficiency and invoke past acceleration events acting on the seed electron population already present in the ICM thermal pool. 

This discovery, only made possible by the unprecedented sensitivity of LOFAR to large angular scales at low frequencies, may hint to a population of low-power, faint and diffuse radio relics, for which re-acceleration has not taken place (or not yet) or is inefficient with respect to standard DSA. This can be explored by the new generation low-frequency arrays (e.g. LOFAR, SKA-low). %

\section*{Acknowledgements}\label{acknowledgments}
\vspace{-0.2cm}
We thank our anonymous reviewer for the helpful scientific feedback.
NTL, KR and FV acknowledge financial support from the ERC  Starting Grant "MAGCOW", no. 714196. AB, CS and EB acknowledge financial support from the ERC  Starting Grant "DRANOEL", no. 714245. NTL thanks Silvia Gandolfi and Raffaele Moretti for extensive support. We thank Dan Wik for useful discussions about IC emission.
MB acknowledges support from the Deutsche Forschungsgemeinschaft under Germany's Excellence Strategy - EXC 2121 "Quantum Universe" - 390833306.
FG and MR acknowledge financial contribution from the agreement ASI-INAF n.2017-14-H.
Radio imaging made use of WSClean v2.6 \citep{2014MNRAS.444..606O} and CASA (https://casa.nrao.edu).
This paper is based (in part) on data obtained with the International LOFAR Telescope (obs. ID LC9\_020, PI F.V) and analysed using LOFAR-IT infrastructure.  
LOFAR (van Haarlem et al. 2013) is the Low Frequency Array designed, constructed by ASTRON and collectively operated by the ILT foundation. 




\bibliographystyle{mnras}
\bibliography{my} 





\bsp	
\label{lastpage}
\end{document}